\documentclass[preprint,12pt,longnamesfirst, epsf]{aastex}

\shorttitle{Environments of Seven Luminous X-Ray Sources}
\shortauthors{Ramsey et al.}

\begin{document}


\title{An Optical Study of Stellar and Interstellar Environments
of Seven Luminous and Ultraluminous X-ray Sources}


\author{
Caitlin J.\ Ramsey\altaffilmark{1},
Rosa M.\ Williams\altaffilmark{2,3},
Robert A.\ Gruendl\altaffilmark{2},
C.-H. Rosie\ Chen\altaffilmark{2},
You-Hua Chu\altaffilmark{2,4},
Q.\ Daniel Wang\altaffilmark{5}}


\altaffiltext{1}{Department of Physics, University of Illinois at 
Urbana-Champaign, 1110 West Green, Urbana, IL 61801}
\altaffiltext{2}{Department of Astronomy, University of Illinois at 
Urbana-Champaign, 1002 West Green Street, Urbana, IL 61801}
\altaffiltext{3}{
Visiting Astronomer, Kitt Peak National Observatory,
National Optical Astronomy Observatories, operated by the
Association of Universities for Research in Astronomy, Inc.\ (AURA)
under a cooperative agreement with the National Science Foundation.}
\altaffiltext{4}{
Visiting Astronomer, Cerro Tololo Inter-American Observatory,
National Optical Astronomy Observatories, operated by the
Association of Universities for Research in Astronomy, Inc.\ (AURA)
under a cooperative agreement with the National Science Foundation.}
\altaffiltext{5}{
Astronomy Department, University of Massachusetts,
B-524 LGRT, Amherst, MA 01003}
\email{cjramsey@uiuc.edu, rosanina@astro.uiuc.edu,
gruendl@astro.uiuc.edu, c-chen@astro.uiuc.edu, chu@astro.uiuc.edu,
wqd@astro.umass.edu}

\begin{abstract}

We have studied the stellar and interstellar environments 
of two luminous X-ray sources and five ultraluminous X-ray 
sources (ULXs) in order to gain insight into their nature.
Archival {\it Hubble Space Telescope} images were used to
identify the optical counterparts of the ULXs Ho IX X-1
and NGC 1313 X-2, and to make photometric measurements 
of the local stellar populations of these and the luminous 
source IC 10 X-1.
We obtained high-dispersion spectroscopic observations 
of the nebulae around these seven sources to search 
for \ion{He}{2} $\lambda$4686 emission and to estimate the 
expansion velocities and kinetic energies of these nebulae.
Our observations did not detect nebular \ion{He}{2}
emission from any source, with the exception of LMC X-1; 
this is either because we missed the \ion{He}{3} regions 
or because the nebulae are too diffuse to produce
\ion{He}{2} surface brightnesses that lie within our 
detection limit.
We compare the observed ionization and kinematics of the 
supershells around the ULXs Ho IX X-1 and NGC 1313 X-2 
with the energy feedback expected from the underlying 
stellar population to assess whether additional energy 
contributions from the ULXs are needed.
In both cases, we find insufficient UV fluxes or 
mechanical energies from the stellar population; thus 
these ULXs may be partially responsible for the ionization 
and energetics of their supershells.  
All seven sources we studied are in young  stellar 
environments and six of them have optical counterparts with 
masses $\gtrsim 7 M_\odot$; thus, these sources are most 
likely high-mass X-ray binaries.  

\end{abstract}


\section{Introduction}

Very luminous X-ray sources with X-ray luminosities ($L_{\rm X}$) 
in the range $10^{38}$--$10^{39}$ ergs s$^{-1}$ are often binaries 
with stellar mass black holes or, less frequently, neutron stars 
as primaries. 
In contrast, ultraluminous X-ray sources (ULXs) with $L_{\rm X} \sim
10^{40}-10^{41}$ ergs s$^{-1}$ are of great interest because their 
bolometric luminosities surpass the Eddington limit for a 100 $M_\odot$
black hole ($1.4\times10^{40}$ ergs~s$^{-1}$). 
Although these unresolved, extragalactic, off-nuclear ultraluminous
sources were first discovered with the {\it Einstein X-ray Observatory} 
\citep{FABBIANO1}, systematic studies of ULXs and similar luminous 
sources have been made possible only recently by the high angular 
resolution afforded by the {\it Chandra X-ray Observatory} 
\citep{HUMPHREY1,SWARTZ1}. 

While some of these objects may be associated with recent supernovae,
or in some cases misidentified background active galactic nuclei 
\citep{SWARTZ1}, there exist ULXs that exhibit variability over short 
time scales, indicating a compact nature \citep[e.g.,][]{STROHMAYER1}.
These compact ULXs may be binary systems containing black holes with 
masses of 10$^2$--10$^4$ $M_\odot$ \citep{COLBERT1,WANG04}.
The existence of such intermediate-mass black holes  (IMBHs) is 
intriguing, as they cannot be produced by current stellar 
evolutionary models.
The hypothesis of a massive accreting black hole nature is 
supported by spectral analyses of several ULXs whose X-ray 
spectra can be successfully fitted with a regular or modified
disk blackbody model, assuming an optically thick accretion 
disk \citep{COLBERT1, MAKISHIMA1, WANG04}.  

It has recently been suggested that thin accretion disks with 
inhomogeneities may produce fluxes exceeding the Eddington 
limit by factors of 10--100 \citep{BEGELMAN1}. Thus 
stellar mass black hole binaries could account for sources with 
luminosities up to $L_{\rm X} \sim 10^{39}$ ergs s$^{-1}$. 
The more luminous sources with $L_{\rm X} \ge 10^{40} \, 
{\rm ergs} \, {\rm s}^{-1}$, however, may still harbor 
IMBHs \citep{MMN2005}.  
Alternatively, if ULXs radiate in the form of anisotropic ``beams"
in the direction of the observer, their luminosities would be 
sub--Eddington, a proposal that is not entirely implausible as 
examples exist of beamed Galactic X-ray sources \citep{LIU2}.  
In this case, ULXs may contain more conventional black holes
or neutron stars with masses $\leq 10 M_{\odot}$ \citep{KING1}. 

The nature of a luminous X-ray source can be ascertained if its 
optical counterpart can be unambiguously identified \citep{LIU3}.
For example, M101 ULX-1 is coincident with a B supergiant, and their 
physical association is confirmed by the unresolved \ion{He}{2} 
$\lambda$4686 emission in the stellar spectrum; thus M101 ULX-1 is 
most likely a high-mass X-ray binary \citep[HMXB;][]{KUNTZ1}.
When optical counterparts cannot be uniquely identified,
the local stellar population can be used to infer the nature of
the sources; for example, the presence of early-type stars supports 
a HMXB origin \citep{LIU3,ROBERTS2,ROBERTS1,
SORIA1}.

Observations of the local interstellar environment can also be 
used to infer the nature of these luminous X-ray sources 
\citep{PM02}.  Luminous X-ray sources are often observed to 
be surrounded by shell nebulae with diameters reaching up to 
several hundred parsecs, called ``supershells"; the X-ray sources 
may contribute to the expansion and ionization of these nebulae
\citep{PAKULL2,MILLER3}.  \citet{P+05} give two potential scenarios 
for the formation and expansion of such nebulae. The first scenario 
posits that the compact  X-ray source and the diffuse nebula were 
both formed in a single supernova explosion; the second, that 
continuous input by stellar winds and/or jets creates a large 
``bubble" around the system.  (Of course, these scenarios are not 
mutually exclusive.)  Calculations of the energy required to create 
a supernova remnant, as required by the first hypothesis, equivalent 
to the observed nebulae around ULXs require explosions of 
$10^{52}-10^{53}$ ergs, which has led some authors to suggest these 
as the remnants of particularly energetic events, called ``hypernovae" 
\citep[e.g.,][]{W2002}.  \citet{P+05} offers the more prosaic
interpretation that the supernova may have exploded in the low-density
surroundings of a pre-existing superbubble, brightening substantially 
as it hit the superbubble wall \citep[e.g.,][]{CM90}. The second
hypothesis, of continuous energy input, encounters a similar problem 
of high input energies, as well as questions about the lifetime over
which such a system will produce strong winds/jets.  Again, these 
problems can be largely ameliorated if the system is within a 
pre-existing cavity.   

Most importantly, nebulae photoionized by X-rays are ``\ion{He}{3} 
regions" in which He is present in the He$^{+2}$ state and its 
recombination leads to \ion{He}{2} $\lambda$4686 emission 
\citep{PAKULL3}. Therefore, the distribution of nebular 
\ion{He}{2} emission and the nebular ionization requirement 
can be used to assess whether the source emits beamed radiation 
\citep[e.g.,][]{PM02,KAARET3}.
Another possibility, as \citet{PAKULL2} suggest, is that these 
nebulae are shock-ionized by same processes that generate the 
mechanical energy inputs discussed above. However, in low-density
regions this process will have very low efficiency.  A third 
scenario is that the nebula was flash-ionized in a very energetic
explosion, and is still in the recombination stage; but the 
lifetime of this phase is not long (of order $10^5$/$n$, where 
$n$ is the ambient density).  All in all, the coupling of 
energetics between the ULX and the nebula is not well understood.
A critical study of supershells associated with luminous X-ray 
sources may provide insight into their nature and energetics, 
just as studies of supershells that are hypernova remnant 
candidates have been used to assess their energetics and confirm 
their nature \citep{LAI1, CHEN1}.

We have obtained high-dispersion echelle spectroscopic 
observations of the nebular environments of seven luminous 
X-ray sources. These data are used to search for \ion{He}{2} 
emission and to study the kinematics of the surrounding medium.  
Additionally, we have utilized archival {\it Hubble Space 
Telescope} ({\it HST}) images to study the local stellar 
populations for some of these sources. In this paper, we report 
our observations and reduction in \S2, and analyze individual 
objects in \S3. The conclusions are summarized in \S4.

\section{Observations and Reduction}

\subsection{High-Dispersion Echelle Spectroscopic Observations}

High-dispersion spectra of the nebular environments of luminous 
X-ray sources were obtained with the echelle spectrographs on 
the 4 m telescopes at Kitt Peak National Observatory (KPNO) 
from an observing run in 2003 November and at Cerro Tololo 
Inter-American Observatory (CTIO) in 2004 January. 
The KPNO observations were all made in the multi-order observing 
configuration, using the 79 line mm$^{-1}$ echelle grating, 
G226-1 cross dispersor, a GG385 filter, the red long focus camera,
and a Tek2K CCD (T2KB). A slit width of 300 $\mu$m (2\arcsec)
was used, giving an instrumental FWHM of 18$\pm$1 km s$^{-1}$ 
in the H$\alpha$ line.  The pixel size of 24 $\mu$m corresponds to 
$\sim$0\farcs26 along the spatial axis and 0.082 \AA\ along the 
dispersion for the H$\alpha$ line.  The observations covered a 
wavelength range of 4500--6900 \AA.

The CTIO observations were made in both the multi-order/short-slit
and the single-order/long-slit observing configurations.
The multi-order configuration used a setup similar to that 
used in the KPNO observations, but with a SITe2K\#6 CCD, which
had the same pixel size and thus the same scales as the T2KB.
These observations covered a wavelength range of 4400--6900 \AA.
The single-order configuration used the 79 line mm$^{-1}$ echelle 
grating with a flat mirror replacing the cross disperser, and a 
post-slit H$\alpha$ filter ($\lambda_c$ = 6563 \AA, $\Delta 
\lambda$ = 75 \AA) was inserted to isolate a single order. 
This spectral coverage is wide enough to include both the 
H$\alpha$ line and the flanking [\ion{N}{2}]~$\lambda\lambda$6548,
6583 lines.  For both configurations, a  slit width of 250 $\mu$m 
(1\farcs64) was used, and the resulting FWHM of the instrumental 
profile was $13.5\pm0.5$ km s$^{-1}$.  

For both the KPNO and CTIO observations, the spectral dispersion 
was calibrated by a Th-Ar lamp exposure taken in the beginning of 
the night, but the absolute wavelength was calibrated against the 
geocoronal H$\alpha$ line present in the nebular observations.
A journal of the echelle observations is given in Table~1.  
A total of seven nebulae were observed.  
The luminous source LMC X-1 is included to assess the feasibility
of our echelle observing configuration, as it is known to possess
a \ion{He}{3} region \citep{PAKULL3}.

As X-ray sources are not visible, target acquisitions often 
need to be made through blind-offset from nearby bright stars.  
The ionized nebulae associated with our target X-ray sources 
are very faint, so their observations are very sensitive 
to sky conditions and moonlight.  
Furthermore, these objects are mostly at megaparsec distances, 
and have angular diameters of only a few arcseconds.  
Owing to these difficulties, we missed the key position of two 
objects, Ho II X-1 and IC 342 X-1 and hence they will be discussed 
in the appendix.  
However, we were able to collect useful H${\alpha}$ spectra for 
Ho IX X-1, IC 10 X-1, M 81 X-6, and NGC 1313 X-2, in addition to
LMC X-1. 
These observations are analyzed in this paper.

\subsection{{\it Hubble Space Telescope} Images}

To study the stellar environments of our sources, we have used the 
archival {\it HST} continuum images of Ho IX X-1, IC 10 X-1, 
M81 X-6, and NGC 1313 X-2.
These observations, listed in Table 2, were made with either
the Wide Field Planetary Camera 2 (WFPC2) or the Advanced 
Camera for Surveys (ACS).
The calibrated pipeline-processed images were acquired from the 
MAST archive and photometry was carried out using the APPHOT 
package in IRAF.


\section{Results for Individual Nebulae}

We first analyze our observations of the \ion{He}{3} nebula 
around LMC X-1 to illustrate the range of capabilities of our 
observing configuration, such as the ability to detect  
\ion{He}{2} emission. 
We use the same method to analyze the nebulae around 
luminous X-ray sources and in some cases, we also analyze 
their stellar populations.  
Below we report on our detailed analysis of the five objects 
for which we obtained useful echelle observations.  
In addition, two objects for which we missed the key positions 
are discussed in the appendix.


\subsection{LMC X-1}

LMC X-1 is a $L_{\rm X}$ = 1--2 $\times 10^{38}$ ergs s$^{-1}$ 
source \citep{SETAL94} in the \ion{H}{2} region N159F which has 
a size of 24$''$, corresponding to 6 pc at the distance to the 
LMC, 50 kpc \citep{Feast99}.  
We have N-S and E-W slit positions centered on the star R148 at 
a position 2\farcs5 east of LMC X-1.  
In each case not only the \ion{He}{2} $\lambda$4686 line but also 
the \ion{He}{2} $\lambda$6560 line was detected (see Fig.\ 1).
The \ion{He}{2} $\lambda$4686 line is narrow, with an observed 
FWHM of 0.39$\pm$0.07 \AA, or 25$\pm$5 km~s$^{-1}$.
The observed FWHM consists of contributions from the thermal
width ($\sim$11 km~s$^{-1}$ for \ion{He}{2} at 10$^4$ K),
instrumental broadening, and turbulence in the gas.
Quadratically subtracting the thermal and instrumental widths
from the observed width, we obtain a turbulent FWHM of
18 km~s$^{-1}$.
The turbulence in the nebula photoionized by LMC X-1 is thus
of order of 10 km~s$^{-1}$, comparable to the isothermal sound
velocity of a 10$^4$ K medium.
There is no highly supersonic motion in the \ion{He}{3} nebula
around LMC X-1.  Therefore it does not appear that  LMC X-1 is
injecting a significant amount of mechanical energy into the 
interstellar medium.  
Our echelle observations of the \ion{He}{3} nebula around 
LMC X-1 illustrate that high-dispersion spectroscopy provides 
a powerful way to study such objects.


\subsection{Holmberg IX X-1 (M81 X-9)}

This ULX was first cataloged as M81 X-9 based on {\it Einstein} 
observations \citep{FABBIANO1}; however, it is projected within 
2$'$ of the nucleus of Holmberg IX, corresponding 
to a projected separation of 2 kpc at a distance D $=$ 3.6 Mpc 
\citep{F1994}, so it has also been referred to as Ho IX X-1. 
It is found at a position near the center of an ionized gas 
shell with $V_{\rm hel}$ $\sim 47-52$ km s$^{-1}$ \citep{MILLER3}.
This velocity is similar to the systemic velocity of Holmberg IX,
64 km s$^{-1}$, as opposed to the \emph{negative} velocities 
expected for M 81 at this position due to its rotation \citep{A1996}. 
If the ULX is indeed associated with this ionized gas shell, 
it is likely a member of Holmberg IX; thus, we will call it 
Ho IX X-1 in this paper.

\subsubsection{Optical Counterpart of Ho IX X-1}

To search for an optical counterpart of Ho IX X-1, we have used 
archival {\it Chandra} ACIS-S observations (Seq Num 600406, Obs ID 4751) 
and {\it HST} ACS images listed in Table 2.  
We have used six stars in the USNO B1.0 catalog 
to determine astrometric solutions for the ACS image, resulting 
in an rms accuracy of $\sim0\farcs25$.  
Our initial comparison of {\it Chandra} X-ray images with the
{\it HST} optical images shows the centroid of the Ho IX X-1
source to be offset by $0\farcs5$ from a bright star at
09$^{\rm h}$57$^{\rm m}$53$\rlap{.}{^{\rm s}}$25, 
$+$69$^\circ$03$'$48\farcs3 (J2000.0).
To verify the alignment of the X-ray and optical images, we
searched for other X-ray sources within the ACIS-S B3 field
and their possible optical counterparts.
We found two faint X-ray sources at 
9$^{\rm h}$58$^{\rm m}$06$\rlap{.}{^{\rm s}}$8, 
$+$69$^\circ$04$'$39\farcs3 (J2000.0) and 
9$^{\rm h}$58$^{\rm m}$12$\rlap{.}{^{\rm s}}$1, 
$+$69$^\circ$04$'$57\farcs0 (J2000.0), respectively.
At these X-ray positions, the ACS images show extended objects,
possibly galaxies, at an offset of $\sim$0\farcs5.
Figure 2 shows that the offsets from the X-ray positions to
the nearest optical objects are similar for all three X-ray
sources. 
If we assume that these offsets are caused by pointing 
uncertainties of {\it Chandra} and {\it HST}, Ho IX X-1 would 
be coincident with the aforementioned star within 0\farcs1.  
Therefore, we suggest that this bright star is the optical
counterpart of the ULX Ho IX X-1.

\subsubsection{Kinematics and Ionization of the Supershell
around Ho IX X-1}

The ionized gas shell around Ho IX X-1 has been observed by
\citet{MILLER3} to have strong [\ion{S}{2}] and [\ion{O}{1}]
line emissions that are characteristic of supernova remnants
(SNRs).  However, the shell size of 25$''\times17''$, corresponding 
to $290\times440$ pc, is much larger than those of known SNRs. 
It is possible that this shell is energized by an OB 
association that \citet{MILLER3} describes as a knot of blue stars.  
The high-resolution H$\alpha$ image presented by \citet{PAKULL2} 
shows a complex nebular morphology consisting of a bright, thin, 
inner ring and a fainter and broader outer ring (see Fig.\ 3a).

We obtained echelle observations for two slit positions
on the supershell around Ho IX X-1: a N-S slit along the eastern
rim and a E-W slit from the eastern rim to the center of 
the shell (see Fig.\ 3).
Nebular emission is detected in H$\alpha$, H$\beta$,
[\ion{O}{3}] $\lambda\lambda$4959, 5007, and [\ion{N}{2}]
$\lambda\lambda$6548, 6583 lines, but not \ion{He}{2} 
$\lambda$4686 or $\lambda$6560.
The velocity structures appear similar in each of the lines
detected, thus we use the strongest detection, the H$\alpha$ 
line, for our analysis of the shell kinematics.
Both slit positions show a bright emission component 
centered at $V_{\rm hel} \simeq 58\pm2$ km~s$^{-1}$.
This velocity is in good agreement with the velocity of 
Holmberg IX, and is adopted as the systemic velocity of 
the shell.
The E-W slit position shows faint emission extending up to
$\Delta V$ = +60 and $-$80 km~s$^{-1}$ from the systemic
velocity, while the N-S slit position shows faint emission
up to $\Delta V$ = +70 and $-$100 km~s$^{-1}$.
The velocity structure does not resemble that of a simple 
expansion: the highest velocity offsets along the N-S slit 
occur at the outer edge of the shell; the E-W slit
shows a brighter blue-shifted component at the inner ring, but
it is dominated by a red-shifted component at the outer ring.  
The 3-D structure of the shell may be pear-shaped, with the inner 
ring corresponding to the top of the pear which is expanding 
toward us (with a negative velocity), and the outer ring
corresponding to the base which is expanding away from us
(with positive velocity), suggestive of a bipolar expansion 
structure.
While the expansion structure cannot be determined unambiguously, 
the overall expansion velocity is likely on the order of 
80--100 km~s$^{-1}$.

The H$\alpha$ flux of the ionized gas shell has been 
reported by \citet{MILLER2} to be $6.4\times10^{-14}$ 
ergs s$^{-1}$ cm$^{-2}$. 
This H$\alpha$ emission requires an ionizing flux of 
$7.3\times10^{49}$ photons s$^{-1}$, if the nebula is
optically thick (to ionizing photons).
Assuming that the emitting material is distributed in a 
prolate ellipsoidal $290\times290\times440$ pc shell with 
a fractional shell thickness ($\Delta R/R$) of 0.1, we 
estimate the rms density of the shell to be 
1.4 H-atom cm$^{-3}$ and the shell mass to be
$6.37\times10^5$ $M_\odot$.
For an expansion velocity of 80 km~s$^{-1}$, the kinetic
energy of the shell is $1.1\times10^{52}$ ergs.

\subsubsection{Local Stellar Population and Energy Consideration}

To determine the roles played by the local stellar population,
and possibly the ULX, in the ionization and energetics of the
supershell, we examine the stellar content using the {\it HST} ACS
images of Ho IX X-1. 
We have carried out photometry of stars in the F450W, F555W, and
F814W bands, and transformed the results into the VEGAMAG system,
which is similar to the Johnson $BVI$ system.
We have produced color-magnitude diagrams (CMDs) in $M_V$ versus
$(B-V)$ and $M_V$ versus $(V-I)$, correcting for only a 
distance modulus of 27.78.
The former was found to be more useful because the stars of 
interest are blue.
This CMD of $M_V$ versus $(B-V)$ for stars within the shell is
presented in Figure 4 (in filled symbols), overplotted with
evolutionary tracks for massive stars from \citet{LS01}.
The highest concentration of stars is distributed near 
$(B-V) \sim -0.1$ and most likely represents the upper part
of the main sequence.
If these are O-stars with $(B-V) = -0.32$, the amount of 
reddening would be $E(B-V) \sim 0.2$.
We have applied this reddening correction to all stars and
plotted them in open symbols in Figure 4.
It is apparent that the main sequence stars detected are within
the mass range of $5$--$40~M_{\odot}$, suggesting an age
of 4--6 Myr for this OB association.

To determine the amount of stellar energy that has been injected
into the surrounding medium, we need to know the number of
supernovae that have exploded in the past and the masses and
temperatures of the most massive living stars, but this
information cannot be determined from the photometry alone.
We therefore assume a Salpeter initial mass function,
and estimate the massive stellar population by extrapolation
from the observed main sequence stars in a lower mass range
that is not seriously affected by incompleteness.
We then use the relation $N = \int^{M_{u}}_{M_{l}}
K M^{-2.35} dM $, where $N$ is the number of stars, $K$ is 
a constant that can be determined from the star counts, $M$ is 
the mass, and $M_{\rm u}$ and $M_l$ are the upper and lower
mass limits.
Five stars are observed to be in the 12--20 $M_\odot$ range,
thus we estimate $K \lesssim 600$.
The number of stars with mass greater than  $20~M_\odot$ is
$\int^{100~M_\odot}_{20~M_\odot} K M^{-2.35} dM \lesssim 7$, 
adopting an $M_u$ of 100 $M_{\odot}$.
Since we observe only one star with $M > 20~M_\odot$, we 
estimate roughly six supernovae have likely exploded.
Assuming each supernova releases $\sim 10^{51}$ ergs of
explosion energy, the total supernova energy input is
approximately $6 \times 10^{51}$ ergs, roughly half of the
observed kinetic energy of the expanding shell.
Additionally, using Starburst99 \citep{Letal99}, we estimate
a collective ionizing luminosity of 0.4--2.0$\times10^{49}$
photons s$^{-1}$, only 1/2 of that required
by the observed H$\alpha$ luminosity of the shell.
Therefore, we conclude that the stellar population alone
cannot provide sufficient energy to produce the observed
kinematics and ionization of the nebula.
It is possible that Ho IX X-1 supplies the additional 
required energy.  
Our echelle spectroscopic observations suggest that there 
may be a bipolar expansion.  
If this is the case, the source may be ``beaming" X-ray 
emission into the nebula toward and away from us.  
As our echelle observations had limited spatial coverage,
the suggested bipolar expansion needs to be confirmed in the 
future with kinematic observations over the full extent of 
the nebula.


\subsection{IC 10 X-1}

Using {\it ROSAT} HRI, IC 10 X-1 was discovered as a rather 
luminous X-ray source within the optical extent of the 
galaxy \citep{B1997}. Recent {\it Chandra} and XMM-Newton 
observations show that
IC 10 X-1 has a mean 0.3-8.0 keV luminosity of $1.2\times10^{38}$ 
ergs 
s$^{-1}$ and shows a large variation by a factor of up to $\sim$6 
on time-scales of
$\sim10^4$ s \citep{Wetal05}.  
It was found to be within $\sim8''$ of the centroid of a 
nonthermal radio supershell 45$''$ in diameter, corresponding 
to 150 pc at the distance of IC 10, $~0.7~$Mpc \citep{Y1993}.

\subsubsection{Optical Counterpart and Stellar Environment}

A possible optical counterpart, an emission-line source,  was 
identified as WR star [MAC92] 17 \citep{M1992}. This was later 
resolved into two components, [MAC92] 17A and B, of which only 
the A component is a WR star \citep{CROWTHER1}.  
Recent {\it Chandra} and {\it HST} observations by \citet{BAUER1} 
with improved spatial resolution and astrometric accuracy
allowed further characterization of the nature of the X-ray source.  
They give a J2000.0 position 
00$^{\rm h}$20$^{\rm m}$29$\rlap{.}{^{\rm s}}$09,
$+$59$^\circ$16$'$51\farcs95 for the point source, and report 
a 0.5--8.0 keV absorbed flux of 
$1.7\times10^{-12}$ ergs cm$^{-2}$ s$^{-1}$ with faint emission 
extending up to $\sim22''$ away.  Their observations support 
the association with [MAC92] 17A, located within 0\farcs23 
of the X-ray source.  The high luminosity and variability of 
this X-ray source, and its likely association with the nearby 
WR star, lead them to hypothesize that the source is most 
likely a massive BH binary with a progenitor that evolved more 
rapidly, and would thus have been more massive than [MAC92] 17A.   

We have carried out photometry using the {\it HST} ACS images 
in the F435W, F606W, and F814W bands, and constructed CMDs
for a distance modulus of 24.23.
The CMD of $M_V$ versus $(B-V)$ for stars in the vicinity of
IC 10 X-1 is presented in Figure 5a.
As IC 10 is located at a low Galactic latitude, the foreground
extinction due to the Galactic plane is evident.
Assuming that the bluest stars are main sequence stars with an 
intrinsic color of $(B-V) = -0.32$, the reddening is $E(B-V) 
\sim 0.85$.
We have applied this reddening correction to all stars and
produced another CMD in Figure 5b, along with evolutionary 
tracks for massive stars \citep{LS01}.
The most massive stars in the vicinity of the IC 10 X-1 have
masses of 20$\pm5~M_{\odot}$.
The lack of more massive stars suggests that this stellar
population was formed about 4--6 Myr ago.
This age is consistent with the what would be expected from 
the presence of a WR star, the optical counterpart of IC 10 X-1.

\subsubsection{Complex Interstellar Environment}

We obtained N-S and E-W echelle spectra with slit positions 
centered near IC 10 X-1 using a slit length of 15$''$ 
(see Fig.\ 6).  No \ion{He}{2} line was detected.
The spectral lines from the E-W slit position were uniform and 
symmetric and thus were useful in estimating a systemic velocity 
for the nebula, $-$330 km s$^{-1}$, close to the systemic velocity 
of IC 10, $-$344 km s$^{-1}$ \citep{dV1991}.  
We apply this systemic velocity to the N-S spectrum which shows 
an expansion structure with low velocity at the center of the 
slit and higher velocities at the edges.  We find an overall 
expansion velocity $\sim80$ km s$^{-1}$, somewhat higher than 
the 50--70 km s$^{-1}$ found by \citet{BR02}.
It is not unusual that high-dispersion spectroscopy can
detect emission at higher velocities than lower-dispersion
spectroscopy.

The velocity structure along the N-S slit position does not
represent a regular expanding shell.
This is understandable as the H$\alpha$ image in Figure 6b 
shows an irregular, filamentary structure that is best described
as a turbulent, frothy interstellar medium.
This interstellar structure may have resulted from the 
explosion of the most massive stars in the population
around IC 10 X-1.
This interstellar environment is too complex to assign 
specific features that have been energized by IC 10 X-1.  
It is thus difficult to assess unambiguously the influence 
of this X-ray source on its surrounding medium.


\subsection{M81 X-6 (NGC 3031 X-11)}

The ULX M 81 X-6 is located near a nebula with an enhanced
[\ion{S}{2}]/H$\alpha$ ratio that was identified as the 
SNR MF 22 \citep{MATONICK1}. 
\citet{PAKULL2} presented a high-quality H$\alpha$ image of this 
region revealing a large 260$\times$350 pc shell structure with 
the SNRs MF 22 and MF 23 on the southern and northern portions 
of the shell, respectively.  
Comparing {\it Chandra} and {\it HST} images, \citet{LIU2} 
identified a unique optical counterpart for M81 X-6, and 
further suggested it to be an O8 V star based on its colors
and magnitudes. 

To show the distribution of stars and the ionized interstellar
gas near the ULX M81 X-6, we present an archival {\it HST} F555W
image in Figure 7a and the continuum-subtracted H$\alpha$ image
from \citet{PAKULL2} in Figure 7b.
The continuum image shows regions of high extinction, and 
the H$\alpha$ emission appears brightest in regions of lower
extinction.
We obtained two echelle spectra with N-S and E-W orientations.
The slit positions are marked in Figure 7b and the data are
shown in Figures 7c and 7d.
The echellograms show a main velocity component at $V_{\rm hel}
= -172$ km~s$^{-1}$ with a FWHM of 30--40 km~s$^{-1}$.
In addition, the N-S slit clearly shows a velocity spike extending
over 250 km~s$^{-1}$, which is typically seen in unresolved 
extragalactic SNRs \citep[e.g.,][]{Detal00}.  The E-W slit shows 
bright emission at the west end, and diffuse emission fanning 
out with high velocity 
at the east end.  Reconstruction of the exact slit positions 
reveals that the spike corresponds to MF 22, confirming 
its SNR nature.    

If the nebulae surrounding the two objects were physically 
related as one supershell, we would expect to see a symmetric 
expansion structure throughout the length of the N-S slit with two 
distinct velocity components, red- and blue-shifted.  We did not, 
however, observe a split-line structure even at the center of the 
``supershell'' that  would indicate an expansion.  The apparent 
shell morphology results from foreground extinction.  We cannot draw 
any conclusions about the role of the ULX in the ionization or 
energetics of the surrounding ISM.


\subsection{NGC 1313 X-2}

NGC 1313 X-2, located $\sim6'$ south of the nucleus of NGC 1313,
was among the first ULXs discovered \citep{FT87}.
The position of this ULX was accurately determined by
\emph{Chandra} observations to be 
3$^{\rm h}$18$^{\rm m}$22$\rlap{.}{^{\rm s}}$18,
$-$66$^\circ$36$'$03\farcs3 (J2000.0), which aided in the 
identification of an $R = 21.6$ pointlike object as the 
optical counterpart of the ULX.
Based on this optical identification and X-ray spectral 
properties, \citet{Z2004} suggest that NGC 1313 X-2 is a black hole 
X-ray binary with a 15--20 $M_\odot$ companion. 
The black hole mass has been estimated to be $\sim$800 $M_\odot$ 
based on the spectral analysis of XMM-Newton observations of
NGC 1313 X-2 \citep{M2003, WANG04}.

NGC 1313 X-2 does not reside in a region of active star formation.
A deep H$\alpha$ image reveals an elongated $25''\times17''$ 
supershell around the ULX, and spectroscopic observations show strong 
[\ion{S}{2}] and [\ion{O}{1}] lines with a FWHM of 80 km~s$^{-1}$ 
centered near the systemic velocity of NGC~1313 \citep{PAKULL2}.
The nebular spectra change abruptly across the supershell, with the 
west side brighter in H$\alpha$ emission and the east side higher 
in [\ion{O}{3}]/H$\alpha$ ratio \citep{Z2004}.

\subsubsection{Optical Counterpart and Stellar Environment}

To examine the optical counterpart and stellar environment
of NGC 1313 X-2, we have used archival {\emph HST} ACS images
in the F435W, F555W, and F814W bands.
The exposure times and observation dates of these ACS images
are given in Table 2.
Figure 8 shows the F435W and F814W images along with the
H$\alpha$ image from \citet{PM02}.
We have used stars in the USNO\,B1.0 catalog to determine 
astrometric solutions for these ACS images, resulting in
an rms accuracy of $\sim$0\farcs5.
With our refined ACS coordinates, it is easy to identify 
the optical counterpart of the ULX, as marked on the ACS images
in Figure 8.
It is also evident that within the H$\alpha$ shell, a higher 
concentration of bright stars exists to the west of the ULX.
This optical identification is consistent with the  
identification of \citet{P+05} based on \ion{He}{2} emission
\citep{MZ+05}. 

We have carried out photometric measurements for the optical
counterpart of the ULX and the bright stars to its west.
The results are transformed to $BVI$ magnitudes and used 
to construct CMDs in $M_V$ versus $(B-V)$ and in
$M_V$ versus $(V-I)$.
We have adopted a distance of 4.1 Mpc \citep{Metal02}, but
did not make any extinction correction.
Figure 9 shows the $M_V$ vs (B-V) CMD plotted with
 evolutionary 
tracks of stars of various masses 
\citep[from][]{LS01}.
The stars are marked in the F555W image in Figure 8b.  

The optical counterpart of NGC 1313 X-2 has $M_V = -3.96\pm0.02$,
$(B-V) = -0.14\pm0.03$, and $(V-I) = -0.11\pm0.04$, equivalent 
to a B1--B2 giant reddened by $E(B-V) \sim 0.1$.
The mass of the star would be $7\pm1~M_\odot$.
The second set of F555W images taken three months after the 
first show a $-0.13\pm0.03$ mag variation in $M_V$.
This variation is small but real.
It is interesting to note that the optical counterparts of
M101 ULX-1 and the ULX NGC 5204 are B supergiants, and 
in the case of M101 
ULX-1 the B supergiant shows no detectable photometric
variation over a timespan of 13 months \citep{LIU3,KUNTZ1}.

The optical counterpart of NGC 1313 X-2 is one of the three 
brightest blue stars in this region; these three blue stars 
are all 7--9 $M_\odot$ early B giants.
The fainter blue stars are most likely main sequence stars 
with masses $\lesssim 10~M_\odot$.
The brightest red star has colors and magnitudes consistent 
with either a 20 $M_\odot$ red supergiant in NGC 1313 or a 
Galactic M8 dwarf at a distance of $\sim$160 pc.
As there are no blue main sequence stars above 10 $M_\odot$
in this region of NGC 1313, we consider the Galactic M dwarf 
a more likely interpretation for this red star. The stellar
 population within the supershell is thus at least 
10$^7$ yr old and consequently no longer contains O stars.

\subsubsection{Ionization and Energetics of the Supershell 
surrounding NGC1313 X-2 }

The 500 $\times$ 340 pc H$\alpha$ shell surrounding NGC 1313 X-2, 
if photoionized
and optically thick, requires an ionizing flux of 
$6.3\times10^{49}$ photons s$^{-1}$, assuming a prolate 
ellipsoidal shell with a fractional shell thickness of 
$\Delta R/R$ = 0.1 and a density of 1 H-atom cm$^{-3}$.
This ionizing flux exceeds that produced by the observed
blue stars, mostly B stars, by two orders of magnitude.
\citet{PM02} suggested that the shell is ionized by
shocks.

To study the energetics of the supershell around 
NGC 1313 X-2, we obtained echelle observations for two slit 
positions (Figure 8).  The N-S slit position was observed in 
the multi-order mode; broad nebular emission is detected in 
H$\alpha$, but the S/N ratio is too low for accurate
velocity measurements.  The \ion{He}{2} line was not detected.  
The E-W slit position was observed in the single-order,
long-slit mode, and this observation has provided the most
useful kinematic information.
The H$\alpha$ line shows a narrow component at a nearly
constant heliocentric velocity of $V_{\rm hel} \sim +380$
km~s$^{-1}$ within the shell boundary and at 30$''$ west
of the shell.
This component represents the local interstellar medium, and
its velocity will be adopted as the systemic velocity of
the shell around NGC 1313 X-2.
The H$\alpha$ line also shows curved blue-shifted and
red-shifted components indicating an expanding shell.
The red-shifted component shows an extreme velocity
offset of +110 km~s$^{-1}$ from the systemic velocity.
The blue-shifted component is blended with the bright 
telluric OH lines at 6571.383 and 6571.386 \AA, so it
is difficult to determine its extreme velocity offset,
but it is at least $-100$ km~s$^{-1}$ from the systemic
velocity.
We thus conservatively assign an expansion velocity of
100 km~s$^{-1}$ to the large shell around NGC 1313 X-2.
This expansion velocity is higher than the 80 km~s$^{-1}$
determined by \citet{PM02} using a lower-resolution 
spectrum.
For a shell $\sim500$ pc in size, this 100 km~s$^{-1}$
expansion velocity is unusually high, and supports the
shock ionization of the nebula, as suggested by 
\citet{PAKULL2}.

The kinetic energy of this shell, assuming the  
aforementioned shell density and geometry, is 
$2\times10^{52}$ ergs.
This energy is much higher than the canonical supernova
explosion energy of 10$^{51}$ ergs and the shell size is
much larger than those of known SNRs.
There could be a large number of normal supernova
explosions that power the expansion of this large supershell,
but this is not supported by the small number of main 
sequence B stars.
It is most likely that the energetic process producing 
the ULX NGC 1313 X-2 is also responsible for powering
this large expanding shell.

\section{Conclusions}

The luminous X-ray sources for which we obtained echelle observations 
were selected because they were surrounded by gaseous nebulae.
It is thus not surprising that all seven of our program X-ray sources are 
in young stellar environments.
For example, the CMDs of stellar populations near Ho IX X-1,
IC 10 X-1, and NGC 1313 X-2 all show main sequence stars typically 
in the mass range 10--25 $M_\odot$. 
The young stellar environment suggests that these X-ray sources are likely HMXBs.
The most direct way to assess the nature of a luminous X-ray source is to identify an 
optical counterpart.  Optical counterparts have been identified for 
six of our program X-ray sources, and in all cases the counterparts are massive 
stars ranging from $\sim7~M_\odot$ to several tens of $M_\odot$.  
These results further support the suggestion that these luminous or 
ultraluminous X-ray sources are HMXBs.  
(See Table 3 for a summary of the stellar and interstellar environments
of the luminous X-ray sources we studied.)
To confirm the association between these stars and the X-ray sources, 
spectroscopic observations of the optical counterparts are needed to 
search for irradiated stellar \ion{He}{2} $\lambda$4686 emission 
\citep[e.g.,][]{KUNTZ1}.  
Once confirmed, spectroscopic monitoring observations can be used 
to measure the orbital parameters of these systems in order to 
determine conclusively whether the ULXs are associated with IMBHs.

The detection of a spatially resolved \ion{He}{2} $\lambda$4686 
emitting nebula around a luminous X-ray source can be used to 
determine whether the X-ray emission is beamed or isotropic.  
However, besides LMC X-1, none of the nebulae around luminous X-ray 
sources for which we obtained clear, 
accurately positioned echelle spectra emitted the \ion{He}{2} 
$\lambda$4686 line. 
To assess our negative results, we compare the nebulae for which 
we did not detect the \ion {He}{2} line to the \ion{He}{2}-emitting 
nebulae associated with LMC X-1 and Ho II X-1.  LMC X-1 is 
surrounded by a dense \ion{H}{2} region with a diameter of 6 pc, 
while Ho II X-1 is in an \ion{H}{2} region 30$\times$45 pc in size.
This size difference reflects the two orders of magnitude 
difference in X-ray luminosity between these two objects. 
The four nebulae for which we do not detect the \ion{He}{2} 
line are large supershells or diffuse \ion{H}{2} regions with 
low concentrations of gas in the vicinity of the X-ray source.  
For such distributions of interstellar gas, X-ray emission from 
the source would be dispersed into a large volume of low-density gas,
resulting in an extremely low surface brightness of \ion{He}{2} 
recombination emission.  Thus the \ion{He}{2} $\lambda$4686 line 
would be difficult, if not impossible, to detect for these regions. 

By comparing the total ionization energy requirements and expansion
velocity of a shell nebula with contributions from the local stellar 
population, we make estimates of energy contributions from Ho IX X-1 
and NGC 1313 X-2.  In both cases, the stellar populations were 
insufficient in producing the energy required for the observed
ionization and kinematics of the nebulae, and thus we concluded 
that these ULXs played a significant role in the energetics of their 
respective supershells.

\appendix{}

\section{Ho II X-1}

This ULX was first detected by {\it ROSAT} observations 
(Zezas et al. 1999). Recent observations by \citet{PM02} 
detected a ``foot-shaped'' nebula with \ion{He}{2} emission 
concentrated in the ``heel'' of the Foot nebula.  Further 
{\it Chandra} observations by \citet{KAARET3} pinpointed 
the ULX at 08$^{\rm h}$19$^{\rm m}$28$\rlap{.}{^{\rm s}}$98,
$+$70$^\circ$42$'$19\farcs3 (J2000.0), indeed within the 
``heel.''   The \ion{He}{3} nebula, ionized by the 
$L_{\rm X} \sim 10^{40}$ erg~s$^{-1}$ X-ray source, has an 
intrinsic FWHM of 2\farcs2, corresponding to 34 pc at a 
distance 3.39 Mpc \citep{PM02}. 

Our slit positions unfortunately missed the \ion{He}{3} 
nebula.  However, the echelle spectra centered at 
08$^{\rm h}$19$^{\rm m}$6$\rlap{.}{^{\rm s}}$0,
$+$70$^\circ$42$'$51\farcs0 (J2000.0), corresponding to the 
middle of the ``Foot," reveal an expanding shell structure, 
and show $V_{\rm exp}$ $\sim$ 65--75 km s$^{-1}$.  
The $V_{\rm hel}$ velocity was found to be +145 km s$^{-1}$, 
corresponding well to the $V_{\rm hel}\sim157$ km s$^{-1}$ of 
Ho II.  An {\it HST} 
H$\beta$  image of the whole foot region shows both nebulosity 
and a high concentration of bright stars.  The detection of a 
blue star at the position of the ULX by \citet{KAARET3} 
suggests a possible optical counterpart with $M_V = -5.5$ to $-5.9$, 
and $(B-V) = -0.2$, consistent with a range of spectral types 
from O4V to B3 Ib.

\section{IC 342 X-1}

IC 342 X-1 is associated with a ``tooth-shaped'' nebula 6$''$ 
in diameter, corresponding to 110 pc at a distance 3.9 Mpc
\citep{ROBERTS1}.
This nebula has a high  [\ion{S}{2}]/H$\alpha$ ratio of 1.2 
and a high [\ion{O}{1}]/H$\alpha$ ratio of 0.4 which is 
characteristic of SNRs \citep{PM02}. Spectroscopy by 
\citet{ROBERTS1} confirms the high [\ion{S}{2}]/H$\alpha$ and 
[\ion{O}{1}]/H$\alpha$ ratios, and finds a high electron 
temperature ($<30,000$ K)  from the [\ion{N}{2}] lines and a 
low density ($<40$ cm$^{-3}$) from the [\ion{S}{2}] doublet. 
The presence of [\ion{O}{3}] in two regions within the nebula
further suggests the possibility of X-ray photoionization. 

Only one set of echelle observations, obtained with a 
N-S oriented slit, revealed narrow 
$H{\alpha}$ and [\ion{N}{2}] lines with a sufficient S/N ratio. 
However, reconstruction of our slit position revealed that it 
was centered on a bright nebular object $3\farcs5$ to the east of 
the target.  The line shows that the nebula has a
 $V_{hel}$ = $-43$ km s$^{-1}$.

\acknowledgements

The authors deeply thank Dr. Manfred Pakull for a very careful
reading of the paper, and for his critical comments which helped
us to improve it.

\clearpage

\begin{figure}
\figurenum{1}
\plotone{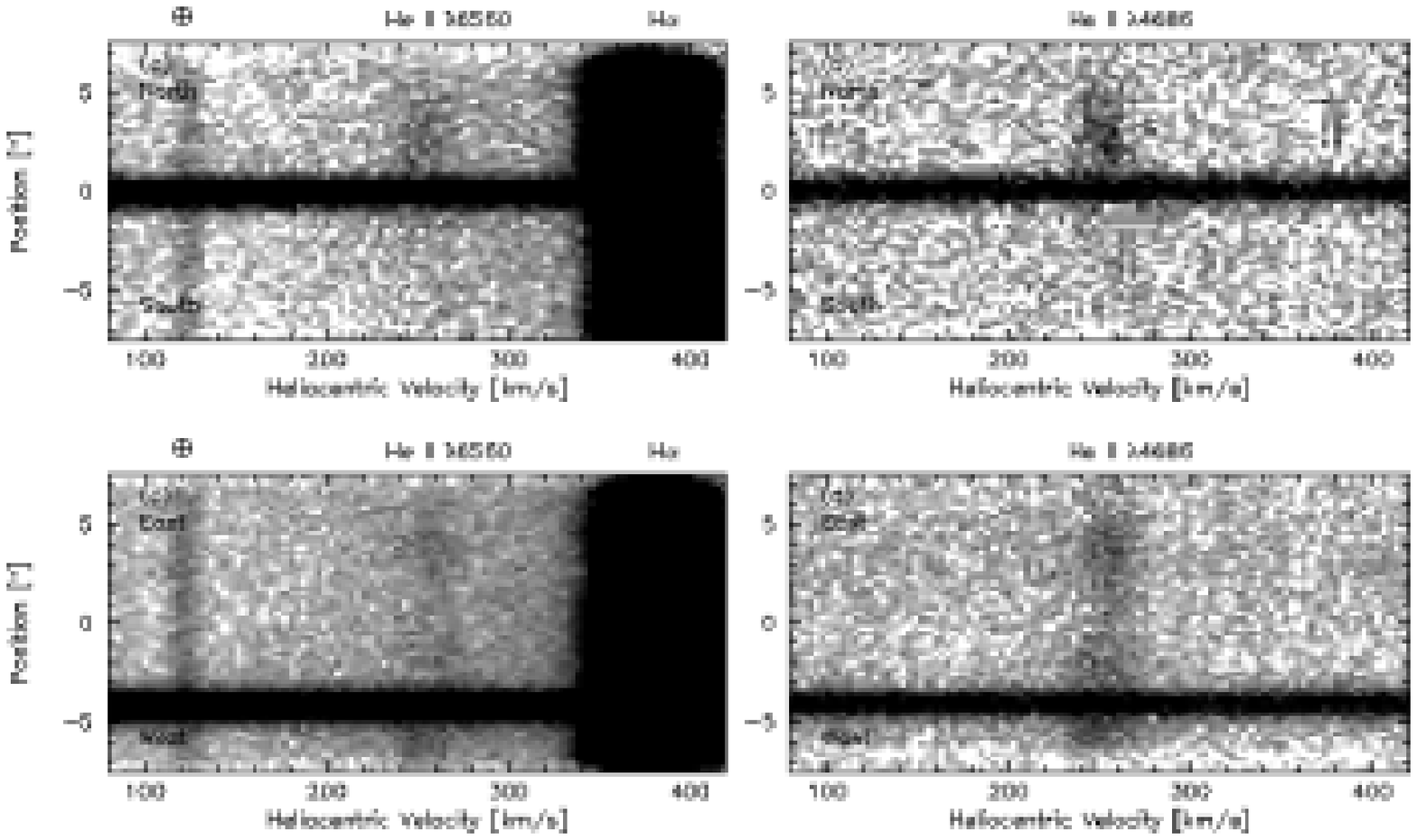}
\caption{Echelle observations of the \ion{He}{3} nebula
around the HMXB LMC X-1.  The \ion{He}{2} $\lambda$6560 
and $\lambda$4686 lines for the N-S slit position are 
shown in (a) and (b).  These \ion{He}{2} lines for the 
E-W slit position are shown in (c) and (d).  The continuum 
source in each echellogram is the star R148 at 2\farcs5 
east of LMC X-1.  The \ion{He}{2} lines are detected to the 
north, east, and west of R148.  The wavelengths are converted 
to heliocentric velocities ($V_{\rm hel}$) of the \ion{He}{2} 
lines.  The bright line appearing at 380 km~s$^{-1}$ is 
the nebular H$\alpha$ at $V_{\rm hel} \sim 260$ km~s$^{-1}$,
and the faint line appearing at 125 km~s$^{-1}$ is the 
geocoronal H$\alpha$ emission.}
\label{fig1}
\end{figure}

\begin{figure}
\figurenum{2}
\plotone{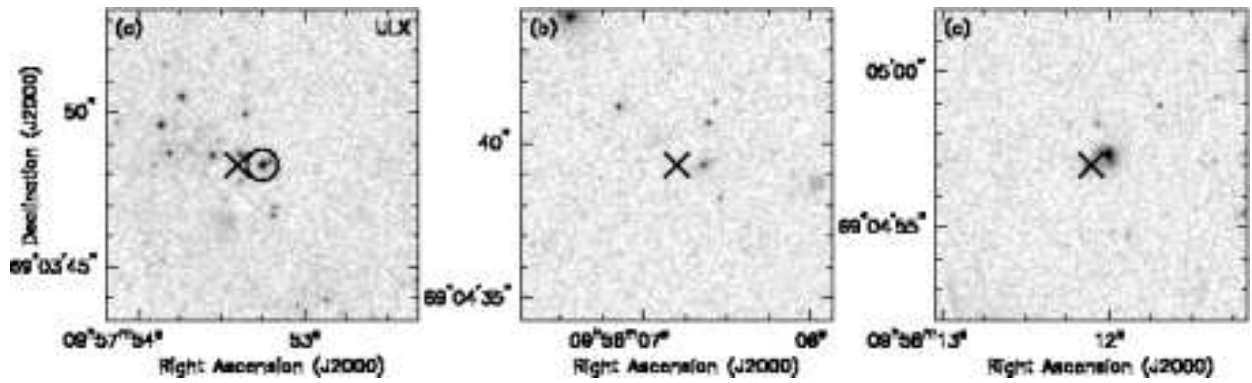}
\caption{Comparison between {\it Chandra} and {\it HST}
pointings.  The centroids of X-ray sources are marked 
by crosses.   The coordinates of the {\it HST} images 
are accurate to within 0\farcs25.  The offsets between
the X-ray sources and their nearest optical objects are
similar and are most likely caused by pointing uncertainties 
of the telescopes.  The star we identify as the optical
counterpart of Ho IX X-1 is marked by a circle in (a).}
\label{fig2}
\end{figure}

\begin{figure}
\figurenum{3}
\plotone{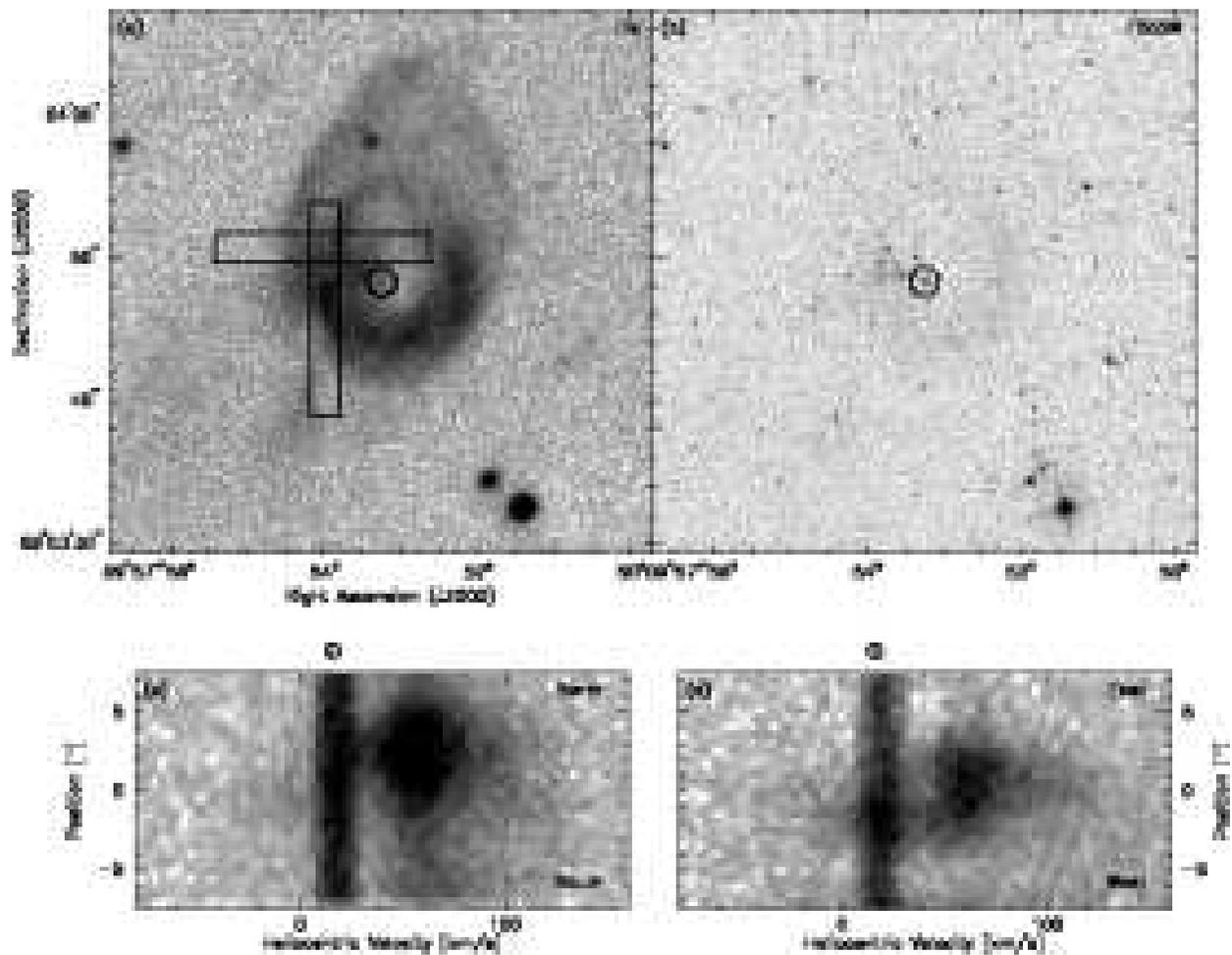}
\caption{Optical images and spectra of Ho IX X-1.  (a) H$\alpha$ 
image taken from \citet{PAKULL2}.  The rectangles mark the echelle 
slit positions.  (b) {\it HST} ACS F555W image with a circle marking 
the position of the optical counterpart of the ULX. (c) echellogram 
of the H$\alpha$ line from the N-S slit position.  (d) same as (c) 
for the E-W slit position.  In each echellogram, the geocoronal 
H$\alpha$ emission is marked by $\oplus$.}  
\label{fig3}
\end{figure}

\begin{figure}
\epsscale{.5}
\figurenum{4}
\plotone{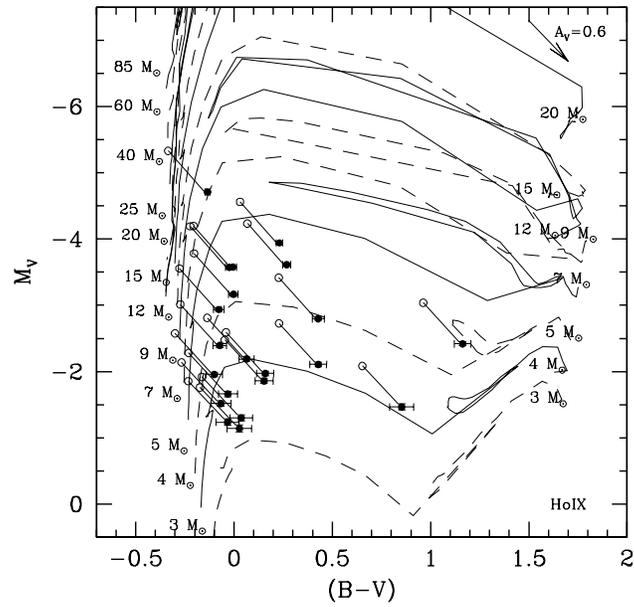}
\epsscale{1}
\caption{Color-magnitude diagram of bright stars within the superbubble 
Ho IX X-1.  A distance modulus of 27.78 was used in deriving the
 absolute magnitude.  The filled and open symbols represent data 
 before extinction correction and with an extinction correction of 
 $A_V=0.6$, respectively.}
\label{fig4}
\end{figure}

\begin{figure}
\figurenum{5}
\plotone{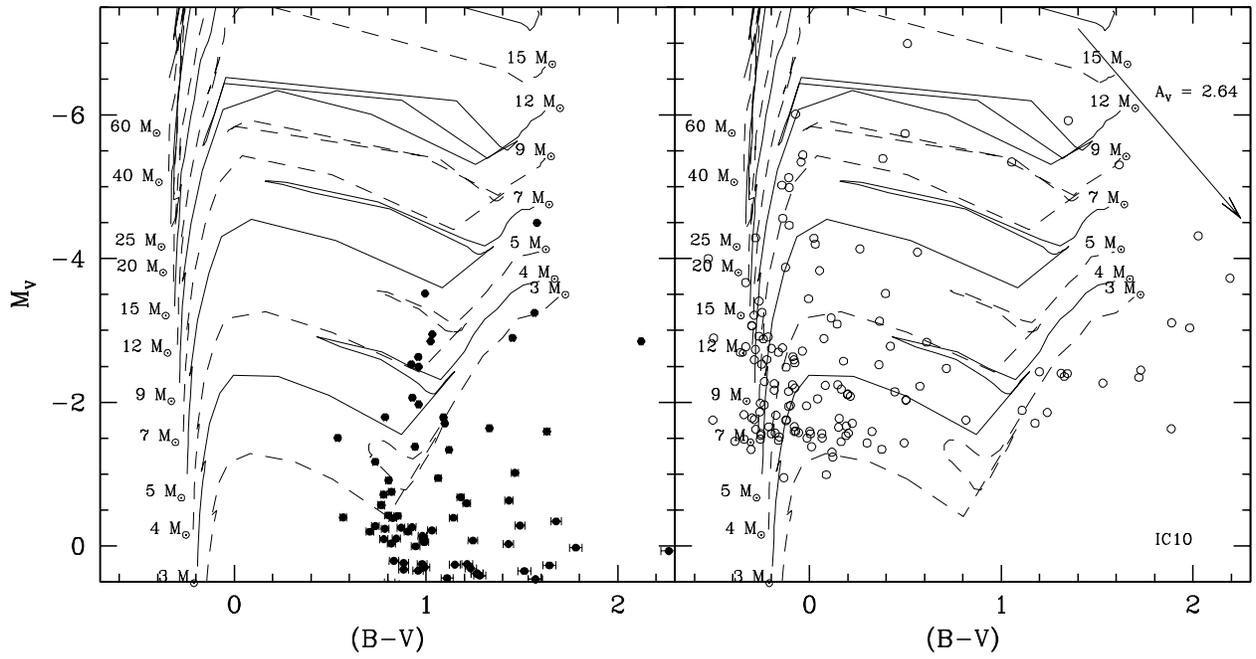}
\caption{Color-magnitude diagram of bright stars near IC 10 X-1.  
Filled symbols in the left panel represent data before extinction 
correction, while open symbols in the right panel represent data 
after an extinction correction $A_V=2.64$.}
\label{fig5}
\end{figure}

\begin{figure}
\figurenum{6}
\plotone{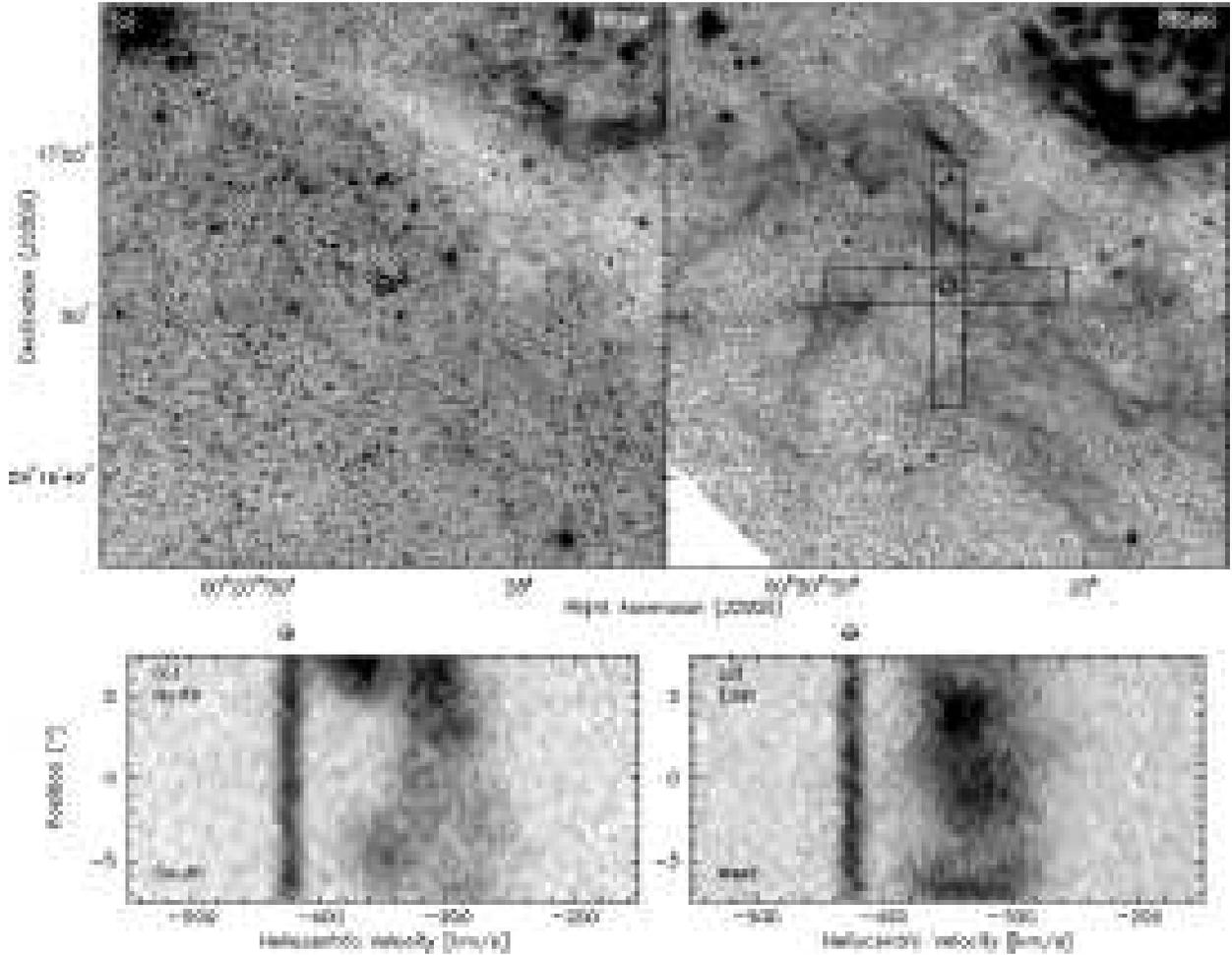}
\caption{Optical images and spectra of the environment of IC 10 X-1.  
(a){\it HST} ACS F606W image of the field around IC 10 X-1. 
(b) Same as (a) in F656N (H$\alpha$) with rectangles marking the 
echelle slit positions and a circle marking the position of the 
optical counterpart of IC 10 X-1.  (c) Echellogram of the H$\alpha$ 
emission for the N-S slit position; the narrow component is the 
telluric OH line at 6553.617 \AA.  
(d) Same as (c) for the E-W slit position.}
\label{fig6}
\end{figure}

\begin{figure}
\figurenum{7}
\plotone{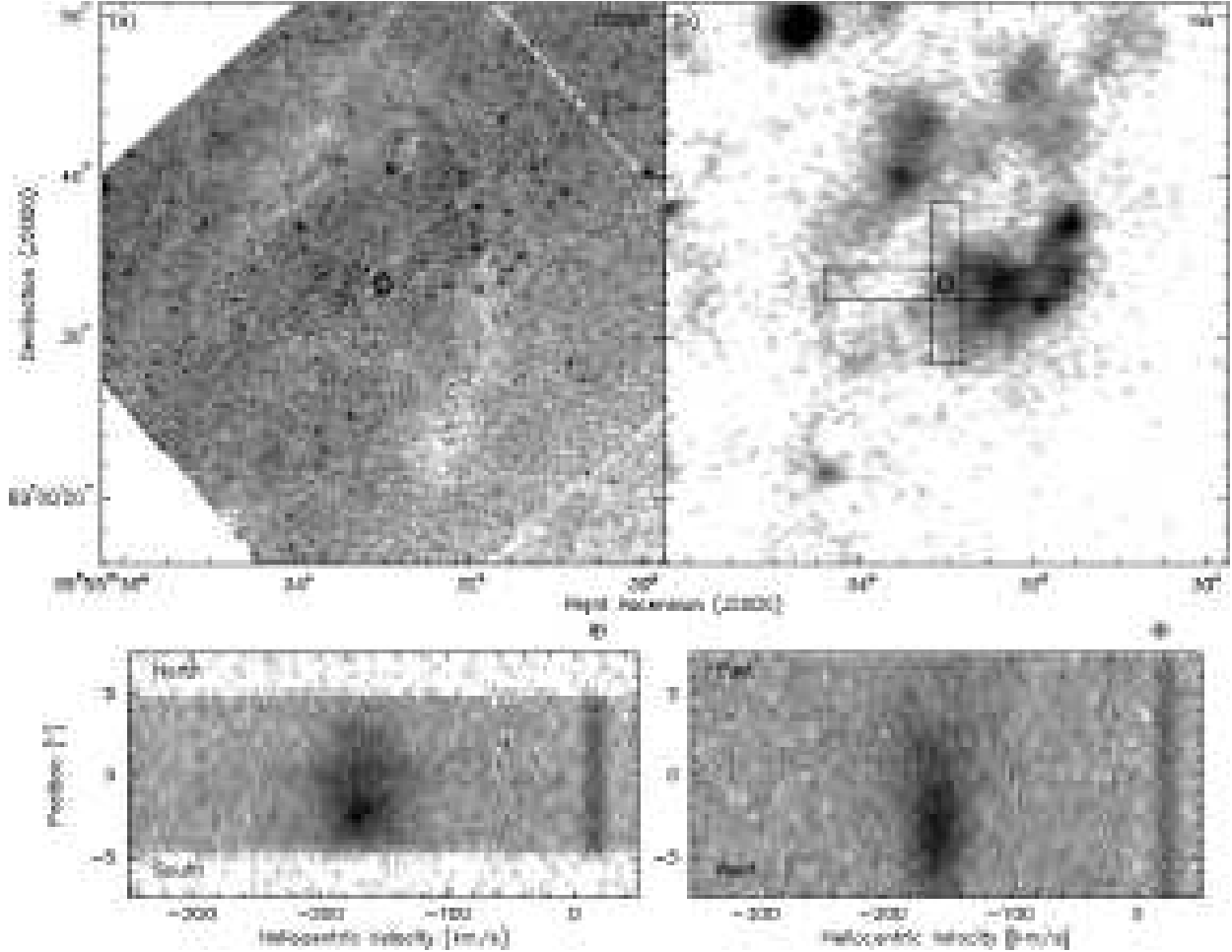}
\caption{Optical images and spectra of the environment of M81 X-6.  
(a){\it HST} WFPC2 F555W image of the field around M81 X-6. 
(b) Continuum-subtracted H$\alpha$ image from \citet{PAKULL2}.  
(c) Echellogram of the H$\alpha$ emission for the N-S slit position;
the narrow component is the geocoronal H$\alpha$ line.  
(d) Same as (c) for the E-W slit position.}
\label{fig7}
\end{figure}

\begin{figure}
\figurenum{8}
\plotone{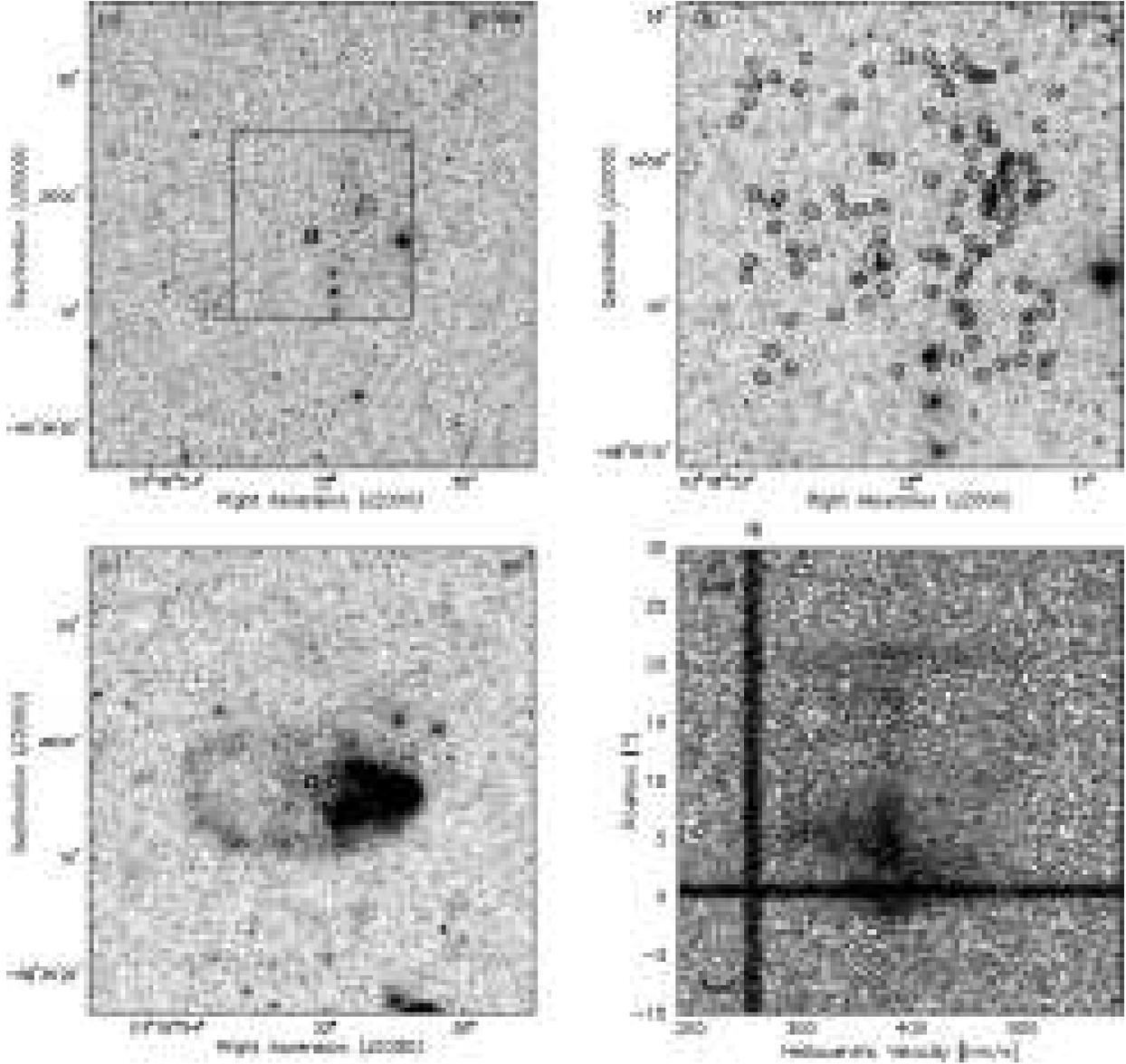}
\caption{Images and spectrum of environment of NGC 1313 X-2. (a) 
{\it HST} ACS image in F555W band with position of the ULX 
marked with a circle.  (b) Enlarged subsection denoted by a 
rectangle in (a).  Stars used in photometric measurements 
are marked with smaller circles.  
(c) H$\alpha$ image from \citet{PAKULL2} showing the supershell 
surrounding NGC 1313 X-2.  (d) Echellogram of the H$\alpha$ emission 
from the E-W slit position; the narrow component is a telluric OH 
line at 6571.383 and 6571.386 \AA.  The continuum source is the 
bright star on the western end of the superbubble shown in (c). }
\label{fig8}
\end{figure}

\begin{figure}
\figurenum{9}
\epsscale{.5}
\plotone{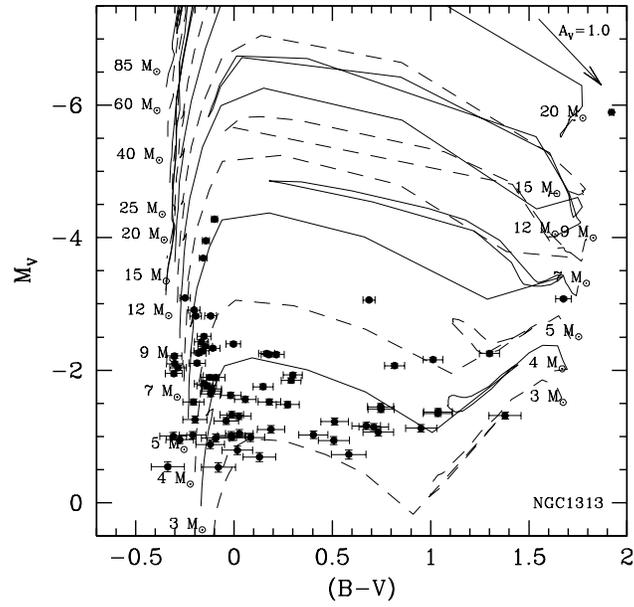}
\epsscale{1}
\caption{Color-magnitude diagram of bright stars within the 
superbubble surrounding NGC 1313 X-2.  No extinction correction 
was made.  A distance modulus of $28.06$ is used in deriving
$M_V$.}
\label{fig9}
\end{figure}

\begin{deluxetable}{rllrlll}
\tablewidth{0pc}
\tablecaption{Journal of Echelle Observations}
\tablehead{
       &  Object & ~Slit~  & Exposure  & ~~Date of~~ &   &   \\
 No.\  &  ~Name~ & Orient. & ~~(s)~~   & Observation & Observatory
 & Remarks 
}
\startdata
 1 & Ho II X-1 & NS& 3$\times$1800 &  2003 Nov 5 & KPNO &    \\
 
 2 & Ho II X-1 & EW& 3$\times$1800 &  2003 Nov 8 & KPNO &    \\
 
 3 &  Ho IX X-1    &  NS  & 4$\times$1800 &  2003 Nov 5 & KPNO & \\

 4 &  Ho IX X-1    &  EW  & 4$\times$1800 &  2003 Nov 7 & KPNO & \\
 
 5 &  IC 10 X-1   &  NS  & 1$\times$1800 &  2003 Nov 5 & KPNO &  Cloudy. \\
 
 6 &  IC 10 X-1   &  NS  & 3$\times$1800 &  2003 Nov 5 & KPNO &\\
 
 7 &  IC 10 X-1   &  EW  & 4$\times$1800 &  2003 Nov 7 & KPNO &\\
 
 8 &  IC 342 X-1  &  NS  & 3$\times$1800 &  2003 Nov 6 & KPNO &\\
 
 9 &  IC 342 X-1  &  NS  & 4$\times$1800 &  2003 Nov 6 & KPNO & Bright sky.\\
 
 10 &  IC 342 X-1  &  EW  & 3$\times$1800 &  2003 Nov 8 & KPNO & \\
 
 11 &  M 81 X-6    &  NS  & 4$\times$1800 &  2003 Nov 6 & KPNO &  \\
 
12 &  M 81 X-6    &  EW  & 4$\times$1800 &  2003 Nov 8 & KPNO & \\

13 & NGC 1313 X-2 &  EW  & 3$\times$1800 &  2004 Jan 8 & CTIO &  Low S/N.\\

14 & NGC 1313 X-2 &  NS  & 3$\times$1800 &  2004 Jan 9 & CTIO & \\

15 & NGC 1313 X-2 &  EW  & 3$\times$1800 & 2004 Jan 12 & CTIO & Single order.\\

16& LMC X-1      &  EW  & 3$\times$1200 &  2004 Jan 8 & CTIO &  \\

17 & LMC X-1      &  NS  & 1$\times$1200 &  2004 Jan 9 & CTIO & \\
\enddata
\end{deluxetable}

\begin{deluxetable}{lllllr}
\tablecaption{Table of Observations}
\tablewidth{0pt}
\tablehead{
\colhead{} & \colhead{} & \colhead{Date of} & 
\colhead{} & \colhead{} & \colhead{Exp. Time}\\
\colhead{Object} & \colhead{PI/PID} & \colhead{Observation} & 
\colhead{Filter} & \colhead{Camera} & \colhead{(s)}}

\startdata

Ho IX X-1 & Miller/9796 & 2004 Feb~~~7& F435W & ACS WFC & $4 \times 630$ \\
          & & & F555W & ACS WFC & $2 \times 580$ \\
          & & & F814W & ACS WFC & $2 \times 580$ \\
IC 10 X-1 & Bauer/9683 & 2002 Oct 12 & F435W & ACS WFC & $6 \times 340$ \\
         & & 2002 Oct 13 & F606W & ACS WFC & $6 \times 360$ \\
         & & & F814W & ACS WFC & $6 \times 360$ \\      
         & Hunter/6406 & 1999 Jun 9 & F656N & WFPC2 & $3\times800$\\  
BM81 X-6 & Bregman/9073 & 2001 Jun 4 & F555W & WFPC2 & $4 \times 500$ \\

NGC\,1313 X-2& Miller/9796 & 2003 Nov 22 & F435W & ACS WFC & $4\times630$ \\
        & &             & F555W & ACS WFC &  $2\times580$ \\
        & &             & F814W & ACS WFC &  $2\times580$ \\
        & & 2004 Nov 22  & F555W & ACS WFC &  $4\times600$ \\

\enddata
\end{deluxetable}

\begin{deluxetable}{llllll}
\tablecolumns{4}
\tablewidth{0pc}
\tablecaption{Summary of Stellar and Interstellar Environments of Luminous X-ray Sources}
\tablehead{
 Object    &  Stellar     &  Optical       & Interstellar    \\
 Name      &  Environment &  Counterpart   &  Environment    }
\startdata
Ho II X-1  & ...        &  BSG   &   \ion{H}{2} region      \\
           &            &        &                         \\
Ho IX X-1  &  blue stars  &  ...   & Supershell, 290$\times$440 pc  \\
           & 4-6 Myr old &        & $V_{\rm exp}$ = 80--100 km s$^{-1}$ \\
           &             &        & $KE = 1.1\times10^{52}$ ergs        \\
           &            &        &                      \\
IC 10 X-1  &  reddened blue stars     & WR star & Interstellar froth  \\
           &           &         & $V_{\rm exp} \sim$ 80 km s$^{-1}$\\
           &            &        &                        \\
IC 342 X-1 &  blue stars  & ...  &   \ion{H}{2} region    \\
           &            &        &                       \\
           &            &        &                        \\
LMC X-1    &  blue stars &  08 III&  \ion {H}{2} region  \\
           &            &        &                        \\
M81 X-6    &  blue stars  &  O8 V  & diffuse \ion{H}{2}  \\
           &            &        & nearby SNR            \\
           &            &        &                       \\
NGC 1313 X-2& blue stars &  B    & Supershell, 496$\times$340 pc  \\
           &       & $\sim7~M_\odot$ &  $V_{\rm exp}$ = 100 km s$^{-1}$  \\
           &             &        & $KE = 2\times10^{52}$ ergs    
\enddata
\end{deluxetable}


\begin{thebibliography}{}

\bibitem[Adler \& Westpfahl(1996)]{A1996}
Adler D.~S., Westpfahl D.~J. \ 1996, \aj, 111, 735

\bibitem[Bauer \& Brandt(2004)]{BAUER1} 
Bauer, F.\ E., \& Brandt, W.\ N.\ 2004,  \apj, 601, L67

\bibitem[Begelman(2002)] {BEGELMAN1}
Begelman, M.\ C.\ 2002, \apj, 568, L97

\bibitem[Brandt et al.(1997)]{B1997}
Brandt, W.~N., Ward, M.~J., Fabian, A.~C., \& Hodge, P.~W.\ 
1997, \mnras, 291, 709 

\bibitem[Bullejos \& Rosado(2002)]{BR02}
Bullejos, A., \& Rosado, M.\ 2002, Revista Mexicana de Astronomia
y Astrofisica Conference Series, 12, 254

\bibitem[Chen et al.(2002)]{CHEN1} 
Chen, C.-H.~R., Chu, Y.-H., Gruendl, R., Lai, S.-P., \& Wang, 
Q.~D.\ 2002, \aj, 123, 2462 

\bibitem[Chu \& Mac Low(1990)]{CM90} Chu, Y.-H., \& Mac Low, M.-M.\ 1990, \apj, 365, 510

\bibitem[Colbert \& Mushotzky(1999)]{COLBERT1} 
Colbert, E.\ J.\ M., \& Mushotzky, R.\ F. 1999, \apj, 519, 89

\bibitem[Crowther et al.(2003)]{CROWTHER1}  
Crowther, P.~A.  et al. \ 2003, \aap, 404, 483

\bibitem[de Vaucouleurs et al.(1991)]{dV1991}
de Vaucouleurs G., de Vaucouleurs A., Corwin J.~R., Buta R.~J., 
Paturel G., \& Fouque P.\ 1991, Third reference catalogue of 
bright galaxies. Springer-Verlag, New York

\bibitem[Dunne et al.(2000)Dunne, Gruendl, \& Chu]{Detal00}
Dunne, B.~C., Gruendl, R.~A., \& Chu, Y.-H.\ 2000, \aj, 119, 1172 

\bibitem[Fabbiano(1989)]{FABBIANO1} 
Fabbiano, G.\ 1989, \araa, 27, 87 

\bibitem[Fabbiano \& Trinchieri(1987)]{FT87} 
Fabbiano, G., \& Trinchieri, G.\ 1987, \apj, 315, 46

\bibitem[Feast(1999)]{Feast99} 
Feast, M.\ 1999, IAU Symp.~190: New Views of the 
Magellanic Clouds, 190, 542 
 
\bibitem[Freedman et al.(1994)]{F1994}
Freedman W.~L., et al.\ 1994, \apj, 427, 628

\bibitem[Georgiev et al.(1991)]{G1991}
Georgiev, Ts.~B., Bikina B.~L., Tikhonov N.~A., \&  Karachentsev I.\
1991, A\&AS, 89, 529

\bibitem[Henkel et al.(1993)]{H1993}
Henkel, C., Stickel M., Salzer J.~J., Hopp U., Brouillet N., 
\& Baudry A.\ 1993, A\&A, 273, L15

\bibitem[Humphrey et al.(2003)]{HUMPHREY1} 
Humphrey, P.~J., Fabbiano, G., Elvis, M., Church, M.~J., \& 
Ba{\l}uci{\' n}ska-Church, M.\ 2003, \mnras, 344, 134 

\bibitem[Kaaret et al.(2004)]{KAARET3}
Kaaret, P., Ward, M.\ J., \& Zezas, A.\ 2004, \mnras, 351, L83

\bibitem [King et al.(2001)] {KING1} 
King, A.\ R., Davies, M.\ B., Ward, M.\ J., Fabbiano, G., 
\& Elvis, M.\ 2001, \apj, 552, L109

\bibitem[Kuntz et al.(2005)]{KUNTZ1} 
Kuntz, K.~D., Gruendl, R.~A., Chu, Y.-H., Chen, C.-H.~R., Still, M., 
Mukai, K., \& Mushotzky, R.~F.\ 2005, \apjl, 620, L31

\bibitem[La Parola et al.(2001)]{LP2001}
La Parola, V., Peres, G., Fabbiano, G., Kim, D.~W., \& Bocchino, 
 F.\ 2001, \apj, 556, 47 

\bibitem[Lai et al.(2001)]{LAI1} 
Lai, S.-P., Chu, Y.-H., Chen, C.-H.~R., Ciardullo, R., \& Grebel, 
E.~K.\ 2001, \apj, 547, 754 

\bibitem[Leitherer et al.(1999)]{Letal99} 
Leitherer, C., et al.\ 1999, \apjs, 123, 3 

\bibitem[Lejeune \& Schaerer(2001)]{LS01} 
Lejeune, T., \& Schaerer, D.\ 2001, \aap, 366, 538 

\bibitem[Liu, Bregman, \& Seitzer(2002)]{LIU2}
Liu,  J.-F. , Bregman, J.\ N., \&  Seitzer, P.\ 2002, \apj, 580, L31

\bibitem[Liu et al.(2004)]{LIU3} 
Liu, J.-F., Bregman, J.~N., \& Seitzer, P.\ 2004, \apj, 602, 249 

\bibitem[Makishima et al.(2000)]{MAKISHIMA1} 
Makishima, K. et al.\ 2000, \apj, 535, 632

\bibitem[Massey et al.(1992)]{M1992} Massey, P., Armandroff, 
T.~E., \& Conti, P.~S.\ 1992, \aj, 103, 1159 

\bibitem[Matonick \& Fesen(1997)]{MATONICK1}
Matonick, D.\ M., \& Fesen, R.\ A.\ 1997, \apjs, 112, 49

\bibitem[M{\' e}ndez et al.(2002)]{Metal02} 
M{\' e}ndez, B., Davis, M., Moustakas, J., Newman, J., Madore, 
B.~F., \& Freedman, W.~L.\ 2002, \aj, 124, 213  

\bibitem[Miller(1995)]{MILLER3} 
Miller, B.\ W.\ 1995, \apj, 446, L75

\bibitem[Miller \& Hodge(1994)]{MILLER2}
Miller,  B.\ W., \&  Hodge, P.\ 1994, \apj, 427, 656

\bibitem[Miller et al.(2005)]{MMN2005}
Miller, N.~A., Mushotzky, R.~F., \& Neff, S.~G. \ 2005, 
\apj, 623, L109

\bibitem[Miller et al.(2003)]{M2003}
Miller, J.~M., Fabbiano, G., Miller, M.~C., \& Fabian, 
A.~C.\ 2003, \apj, 585, L37

\bibitem[Mizuno, Kubota \& Makishima(2001)]{MIZUNO1} 
Mizuno, T., Kubota, A. \& Makishima, K.\ 2001, \apj, 554, 1282

\bibitem[Mucciaelli et al.(2005)]{MZ+05}Mucciaelli, P., Zampieri, 
L., Falomo, R., Turolla, R., \& Treves, A. 2005, astro-ph/0510085

\bibitem[Pakull \& Angebault(1986)]{PAKULL3}
Pakull, M.\ W., \& Angebault, L.\ P.\ 1986, Nature, 322, 511 

\bibitem[Pakull \& Mirioni(2002)]{PM02}  
Pakull, M.\ W., \&  Mirioni, L.\ 2002, in proceedings ``New Visions of the X-ray Universe in the XMM-Newton and Chandra Era", held 26-30 November 2001 in The Netherlands (astro-ph/0202488)

\bibitem[Pakull \& Mirioni(2003)]{PAKULL2}  
Pakull, M.\ W., \&  Mirioni, L.\ 2003,  RevMexAA 
(Serie de Conferencias) 15, 197

\bibitem[Pakull, Grise, \& Motch(2005)]{P+05}Pakull, M.\ W., Grise, F. \& Motch, C.\ 2005, IAU Symp. No. 230, ``Populations of High-Energy Sources in Galaxies", held on 2005 August 15-19 in Dublin, Ireland

\bibitem[Roberts et al.(2003)]{ROBERTS1} 
Roberts, T.~P., Goad, M.~R., Ward, M.~J., \& Warwick, R.~S.\ 2003, 
\mnras, 342, 709 

\bibitem[Roberts et al.(2001)]{ROBERTS2} 
Roberts, T.~P., Goad, M.~R., Ward, M.~J., Warwick, R.~S., O'Brien, 
P.~T., Lira, P., \& Hands, A.~D.~P.\ 2001, \mnras, 325, L7

\bibitem[Schlegel et al.(1994)]{SETAL94}
Schlegel, E.~M., Marshall, F.~E., Mushotzky, R.~F., Smale, A.~P., 
Weaver, K.~A., Serlemitsos, P.~J., Petre, R., \& Jahoda, K.~M.\ 1994,
\apj, 422, 243S 
 
\bibitem[Soria et al.(2005)]{SORIA1} 
Soria, R., Cropper, M., Pakull, M., Mushotzky, R., \& Wu, K.\ 
2005, \mnras, 356, 12 
 
\bibitem[Strohmayer \& Mushotzky(2003)]{STROHMAYER1} 
Strohmayer, T.~E., \& Mushotzky, R.~F.\ 2003, \apjl, 586, L61

\bibitem[Swartz et al.(2004)]{SWARTZ1} 
Swartz, D.\, A., Ghosh, K.\ K., Tennant, A.\ F., \& Wu, K.\ 2004, 
\apjs, 154, 519

\bibitem[Wang(2002)]{W2002}
Wang, Q.~D. \, 2002, \mnras, 332, 764

\bibitem[Wang et al.(2004)]{WANG04} 
Wang, Q.~D., Yao, Y., Fukui, W., Zhang, S.~N., \& Williams, 
R.\ 2004, \apj, 609, 113

\bibitem[Wang et al.(2005)Wang, Whitaker, \& Williams]{Wetal05}
Wang, Q.~D., Whitaker, K.~E., \& Williams, R.\ 2005, \mnras, 698

\bibitem[Yang \& Skillman(1993)]{Y1993} Yang, H., \& 
Skillman, E.~D.\ 1993, \aj, 106, 1448 

\bibitem[Zampieri et al.(2004)]{Z2004} 
Zampieri, L., Mucciarelli, P., Falomo, R., Kaaret, P., 
di Stefano, R., Turolla, R., Chieregato, M., \& Treves, A.\ 
2004, Nuclear Physics B Proceedings Supplements, 132, 387 

\bibitem[Zezas et al.(1999)]{Z1999} Zezas, A.~L., 
Georgantopoulos, I., \& Ward, M.~J.\ 1999, \mnras, 308, 302 

\end{thebibliography}
\end{document}